# Excitation of epsilon-near-zero resonance in ultra-thin indium tin oxide shell embedded nanostructured optical fiber

*Khant Minn[1], Aleksei Anopchenko[1], Jingyi Yang[1], and Ho Wai Howard Lee [1,2] †*

[1]Department of Physics, Baylor University, Waco, TX 76798, United States

[2]The Institute for Quantum Science and Engineering, Texas A&M University, College Station, TX 77843, United States

†Corresponding author e-mail: Howard_Lee@Baylor.edu

**Abstract**

We report a novel optical waveguide design of a hollow step index fiber modified with a thin layer of indium tin oxide (ITO). We show an excitation of highly confined waveguide mode in the proposed fiber near the wavelength where permittivity of ITO approaches zero. Due to the high field confinement within thin ITO shell inside the fiber, the epsilon-near-zero (ENZ) mode can be characterized by a peak in modal loss of the hybrid waveguide. Our results show that such in-fiber excitation of ENZ mode is due to the coupling of the guided core mode to the thin-film ENZ mode. We also show that the phase matching wavelength, where the coupling takes place, varies depending on the refractive index of the constituents inside the central bore of the fiber. These ENZ nanostructured optical fibers have many potential applications, for example, in ENZ nonlinear and magneto-optics, as in-fiber wavelength-dependent filters, and as subwavelength fluid channel for optical and bio-photonic sensing.

Optical response of near-zero refractive index systems has been a topic of interest recently as the electromagnetic field inside the media with near-zero parameters, i.e., vanishing permittivity and permeability values, have been shown to exhibit unique optical properties. Those features may be exploited in various optical applications such as wavefront engineering, radiation pattern tailoring [1,2], non-reciprocal magneto-optical effects [3], nonlinear ultrafast optical switching [4,5], dielectric permittivity sensing [6,7], and broadband perfect absorption [8,9]. Recent studies suggest that epsilon-near-zero (ENZ) properties can also be observed in a single highly doped conducting oxide thin film. Unique properties observed include enhanced absorption in transparent conducting oxide (TCO) ENZ layers [8,10-13], advanced resonant coupling properties with antenna [14-16], and strongly enhanced nonlinear response and light generation in a TCO slab [5,17-20]. In addition, electrical tuning of conducting oxide materials to the ENZ regime results in efficient light manipulation and modulation [21-23]. However, most of the studies on ENZ optical properties are limited to the excitation of ENZ mode in the planar structures or meta-surfaces with short interaction length, restricting the excitation platform for novel optical applications.

Photonic crystal fiber (PCF) or micro-structured optical fiber consists of hollow channels running along the entire length of a glass strand, providing unique platform with long interaction length and engineerable dispersion for the studies of nonlinear optics, optical communication, optical/bio sensing, etc [24]. The optical properties of PCFs can be changed by filling the hollow channels with materials such as semiconductors and metals to excite the Mie resonances and surface plasmon resonances [25,26]. Those metal/semiconductor infiltrated fibers have been proposed to use for optical sensing and in-fiber device applications. Simplified version of photonic crystal fiber with enhanced light-matter interaction could be achieved by introducing a nanoscale hollow channel into conventional optical fiber (e.g., nanobore optical fiber). Such nanobore fiber further allows the light coupling to the plasmonic modes of gold nanowire for polarization conversion [27] and optical detection of virus in nano-fluid channel [28].

In this work, we investigate the ENZ mode excitation in optical fiber platform and demonstrate that propagating fields can be confined inside a region coated with ENZ conducting oxide material which is incorporated into a nanostructured optical fiber waveguide. The novel hybrid optical fibers could be used as a platform for highly sensitive optical sensing and magneto/nonlinear-ENZ optical studies.

The ability to confine electromagnetic energy in a small space at the medium's ENZ wavelength motivates a search for highly confined propagating polariton modes using ENZ materials. It has been experimentally shown that three layer structures, where a sub-wavelength thin layer of indium tin oxide is sandwiched between two dielectric layers, can support ENZ polariton modes [16,29,30]. Here we present the existence of propagating mode in an optical fiber with nano-hollow channel modified with an ENZ layer within which enhanced field is excited. Our proposed ENZ fiber waveguide design is a modified version of nanobore fiber which consists of three concentric cylindrical shells (Fig. 1). The two outermost shells act as cladding and core respectively. The innermost thin shell is made of ENZ material that surrounds the hollow central channel. The sub-wavelength thickness of ENZ shell ensures that the guided core fiber mode can be coupled to the thin film ENZ mode whose domain of existence is limited to film thickness much less than the plasma wavelength of the material at which its permittivity vanishes [12].

Indium-tin-oxide (ITO), a CMOS compatible TCO material, is used as the ENZ medium in the analysis. The real part of ITO permittivity crosses zero at the bulk plasmon resonant wavelength, which can be tuned in near-infrared by controlling the carrier concentration of the material [31]. The frequency dependent complex permittivity of ITO was calculated using the Drude model. (See ITO material and optical properties in *Methods* and Supporting Information). For the designed ITO carrier concentration of $10^{21}$, the real part of permittivity function of the ITO crosses zero at 1068 nm and with small imaginary part of permittivity ($Im(\varepsilon) = 0.41$). This carrier concentration of ITO could be achieved by various deposition techniques, for instance magneton sputtering and atomic layer deposition (ALD) [22,23,31-33]. ALD or wet chemistry techniques could be used to fabricate conducting oxide nano-shell inside the hollow channel of the fiber[33-35].

To understand the coupling mechanism between the guided mode in the nanobore fiber and the thin film ENZ mode, we investigate their dispersion characteristics. The effective index of the fundamental guided core mode of the nanobore fiber as a function of excitation wavelength was modelled using finite difference numerical waveguide analysis method (see *Methods*). The effective index of non-radiating thin film ENZ mode supported by the glass-ITO-air three layer structure was modelled using the transfer matrix method (see *Methods*). The coupling between the fundamental mode of the nanobore fiber and thin film ENZ mode occurs at the phase matching wavelength, at which point the effective index functions of the two modes intersect and their momenta are equal. Our results show that at the phase matching wavelength, the field confinement inside the ITO shell in the ENZ fiber is the highest and the mode has the highest loss, thus confirming the excitation of thin film ENZ mode in the fiber structure.

A schematic of the proposed ENZ optical fiber design is shown in Fig. 1(a). Section (I) of the structure is the hollow nanobore fiber which consists of the outer silica cladding, GeO$_2$ wt. 9% doped silica core of 4 μm diameter and a hollow central channel of 200 nm diameter. Section (II) depicts the ENZ fiber where the inner surface of the central hollow channel is coated with 10nm layer of ITO. Figure 1(b) shows the glass-ITO-air three layer structure that is considered in calculating the thin film ENZ mode. The structure is comprised of a thin ITO layer having the same thickness as the ITO shell inside the ENZ fiber sandwiched between the air and glass half spaces.

The dispersion of fundamental mode of the hollow nanobore fiber was obtained using finite difference numerical waveguide simulations on the cross-section of the nanobore fiber as depicted in the right insert of Fig. 1(a). A frequency dependent real part of the effective index of the mode $n_{eff}$ was calculated ($n_{eff} = c\beta/\omega$, where $\beta$ is the propagation constant in the fiber for a given wavelength). To obtain the dispersion curve of the thin film ENZ mode, we investigated the glass-ITO-air three layer geometry shown in Fig. 1(b) using the transfer matrix method. To ensure the excitation of the non-radiative ENZ mode, the ENZ mode is excited from the glass half space using Kretschmann configuration [36]. The reflectance was calculated for varying incident angles and wavelengths. For small enough thickness of ITO layer ($d < \lambda/50$), light incident from glass onto ITO nanolayer is perfectly absorbed at large angles and at resonant wavelength which corresponds to the ENZ polariton mode [16]. Thus for each incident angle, the wavelength that corresponds to minimum reflectance is traced to calculate the effective index of the thin film ENZ mode (see Fig. S3 in the Supporting Information). The thin film ENZ dispersion curves for 5 nm, 10 nm, 15 nm, and 20 nm ITO layer thicknesses were calculated to show that the phase matching condition with fiber core modes depends on the thickness of the ITO layer.

To understand the phase matching condition, the effective index of the thin film ENZ modes for four different ITO layer thicknesses (dash curves) and the fundamental mode of hollow nanobore fiber (blue dotted curve) were calculated and depicted in Fig. 2. As the ITO thickness increases, the thin film ENZ dispersion curve shifts toward longer wavelengths and thus the phase matching wavelength (crossing wavelength) with the fiber core mode. Thin film dispersion curves intersect the fiber core mode dispersion at 1070 nm, 1079 nm, 1088 nm and 1098 nm wavelengths for 5 nm, 10 nm, 15 nm and 20 nm ITO layer thicknesses respectively. Thus in the ENZ fiber, ENZ mode is expected to be excited at higher wavelengths for thicker ITO shells.

As a next step, we verified the excitation of ENZ modes in the fiber at the above resonant wavelengths. Finite difference method was used to solve the Maxwell's equations on a cross-section of the ENZ fiber waveguide. The resulted fundamental mode has highest spatial field distribution within the ITO shell compared to higher order modes. This ENZ mode was tracked over a wavelength range to calculate the modal loss ($loss = -20 log_{10} e^{-2\pi k/\lambda_0}$, where $k$ is the imaginary part of the effective index). The modal loss curves were calculated in the same way for four different ITO shell thicknesses: 5 nm, 10 nm, 15 nm and 20 nm. Figure 2 shows the modal losses of the fundamental mode of the ENZ fiber for four ITO shell thicknesses (color shaded areas). Peak losses are observed at wavelengths 1071 nm, 1080 nm, 1092 nm and 1102 nm for ITO shell thickness 5 nm, 10 nm, 15 nm and 20 nm respectively. These resonances are in good agreement with the phase matching wavelengths of the nanobore fiber mode and the thin film ENZ modes calculated above. The peak in the loss spectra is resulted from the excitation of ENZ thin film mode and the strong confinement/absorption by the ITO nanolayer.

The slight discrepancy between the phase matching and the peak loss wavelengths can be attributed to the mismatch between the excitation light sources: plane wave in the case of three layer geometry and core guided eigen-mode source in the nanobore fiber. Another contributing factor is the curvature of the ITO shell inside the fiber which is cylindrical in contrast to the flat geometry of three layer structure that is assumed when calculating the thin film ENZ modes. The resonance discrepancy is greater with increasing shell thickness due to the mismatch in the two geometries. To study the nature of ENZ mode in the fiber, we plot the spatial electric field distributions of the fundamental mode supported by the ENZ fiber with 10 nm thick ITO shell at the ENZ wavelength and off-ENZ wavelength (Fig. 3). The field distribution does not exhibit radial symmetry because the excitation fiber core mode is linearly polarized ($HE_{11}$). The radially polarized higher order mode ($TM_{01}$) is not supported by the nanobore fiber at the high wavelength regime (Supporting Information S2). At 1080 nm, which is the ENZ mode phase matching wavelength for 10 nm thick ITO layer, magnitude of electric field is highly confined inside ITO shell (ENZ region) as shown in Fig. 3(b) and (e). The high field confinement in the ENZ nano-layer results in the highest modal loss at this wavelength as seen in Fig. 2 (yellow solid curve). At shorter wavelength (900 nm), the refractive index of ITO resembles that of a dielectric with real part of value (n = 1.022) falls between that of air at the center and doped silica core (n = 1.459) (Supporting Information S1). In this regime, the field distribution resembles the profile of the fundamental mode of hollow nanobore fiber as can be observed in Fig. 3 (a) and (d). At longer wavelength (1300 nm), ITO is in essence metallic-like having negative real part of permittivity ($\varepsilon = -1.707 + 0.732i$) with real part of index (0.274) smaller than imaginary part (1.335). Thus, the field is mostly contained in the core and central air channel, decaying inside the ITO shell as depicted in Fig. 3(c) and (f).

The field enhancement in the thin ITO shell suggests that the modal properties of the fiber ENZ mode may be sensitive to the perturbations of the surrounding medium's optical properties, in particular, the dielectric permittivity of the constituents in the central hollow channel. Thus, we investigated the dependence of the modal loss spectrum on the refractive index of the central channel. The fiber ENZ modal loss spectra calculated for five different fluids placed in the central channel of the fiber is depicted in Fig. 4. The refractive index of the fluid at the resonant wavelength was stated inside the parentheses next to the name of the fluid. The permittivity functions of the fluids are obtained from other reports[37,38]. The thickness of the ITO shell was kept at 20 nm. As seen in the figure, the resonant wavelength shifts from 1102 nm for air to 1156 nm for chloroform. The modal loss for chloroform is found to be 679 dB/cm, which is almost five times greater than with air at the central channel. Increasing the refractive index of the central region in the nanobore fiber effectively shifts the dispersion curve of both the guided fiber core mode and planar thin film ENZ mode (Supporting Information S3). This in turn results in the phase matching wavelength getting larger due to the change of the index of the hollow channel.

Figure 5 shows the comparison between the ENZ mode profiles for air and chloroform at the center of the ENZ fiber. Similar to the analysis in the planar structures, the field confinement of the ENZ mode is stronger with the external medium of chloroform, and thus the larger absorption of light was observed in the ENZ fiber with higher refractive index in the central nano-channel. Similar trend is observed in the field profiles for the three layer geometry (Supporting Information S3).

The strong dependency on the surrounding dielectric can be exploited for novel optical/bio sensing purposes due to the unique feature of the enhanced field confinement of the ENZ mode in the fiber. For wavelength-based sensing, the average refractive index sensitivity of the ENZ fiber, defined as $\Delta\lambda/\Delta N$ where $N$ is the refractive index of the material in the central channel of the fiber is found to be 121 nm/RIU. The observed sensitivity is comparable with other in-fiber sensing device such as Mach–Zehnder interferometer with waist-enlarged fusion bitaper [39,40] and it has better performance than some grating-based sensors [41,42]. In addition, this type of ENZ fiber sensor could be used for sensing material with wide range of refractive index, including refractive index between 1.3-1.4 in which most of the biomaterial lie on. The sensitivity of the ENZ fiber could be significantly enhanced by optimizing the materials and geometry, such as ITO thickness, core/hole diameters, choices of ENZ materials, and the core mode dispersion (e.g. engineered dispersion with additional cladding holes[24,43]), however, it is beyond the scope of this paper.

In summary, we have investigated the ENZ modes supported by a hollow fiber waveguide modified with a thin conducting oxide layer that is capable of confining high intensity fields inside a subwavelength nano-channel. At the dielectric to metal crossover point of the conducting oxide, the permittivity approaches zero enabling the excitation of ENZ mode characterized by high field enhancement in the ENZ region. The sensitivity of the waveguide's modal loss to the central channel refractive index can be exploited to sense refractive index of medium and therefore the ENZ fiber has potential applications in optical/bio-sensing. In addition, due to the excitation of the highly confined ENZ mode in the optical fiber waveguide, the ENZ fiber could be potentially useful in studying nonlinear and magneto-optics as well as enhanced quantum emission near ENZ medium, and transmitting optical energy below the diffraction limits in fiber.

## Methods

Numerical simulations of the waveguide structure were carried out using the MODE Solutions software from Lumerical Solutions, Inc. The effective index of the ENZ mode supported by three layer structures was calculated using of a MATLAB code developed by G. Figliozzi and F. Michelotti at the University of Rome "La Sapienza" (Italy)[44] based on the transfer matrix method [45]. In all simulations, permittivity function of silica is modeled using the Palik data [46]. Permittivity of $GeO_2$ doped silica ($GeO_2$ wt. 9% doped silica) is obtained from the literature [47]. For ITO, the Drude model is used with the parameters: the electron concentration of $1.0 \times 10^{21}$ $cm^{-3}$, $\varepsilon_\infty = 3.6$, $\Gamma = 2.0263 \times 10^{14}$ $s^{-1}$, $\omega_p = 3.3722 \times 10^{15}$ $s^{-1}$, [48] In the index sensing analysis, the permittivity functions of the liquids that are placed inside central channel to be analyzed are obtained from literature [37,38].

## Data Availability

The datasets generated during and/or analyzed during the current study are available from the corresponding author on request.

## Acknowledgments

This work was supported in parts by the Defense Advanced Research Projects Agency (grant number N66001-17-1-4047), the Young Investigator Development Program, and the Vice Provost for Research at Baylor University. The authors acknowledge the support of the usage of Kodiak high performance computing cluster at Baylor University. A.A. acknowledges the Office of the Vice Provost for Research at Baylor University for the postdoctoral research fellowship. A. A. wish to thank Prof. F. Michelotti for the use of the MATLAB code for the transfer matrix calculations.

## Author Contributions

K.M., A.A., H.W.H.L. designed and conceived the project. K.M., A.A., and J.Y. performed numerical simulations. H.W.H.L. supervised the project. All authors wrote the paper, discussed the results, and commented on the manuscript.

The authors declare no competing financial interests.

## Supporting Information

Calculation details of ENZ wavelength of ITO, cut-off wavelengths of higher order modes of ENZ nanobore fiber, simulation details of thin film ENZ mode dispersion and field profiles in planar three layer geometry. This material is available from the corresponding author upon request.

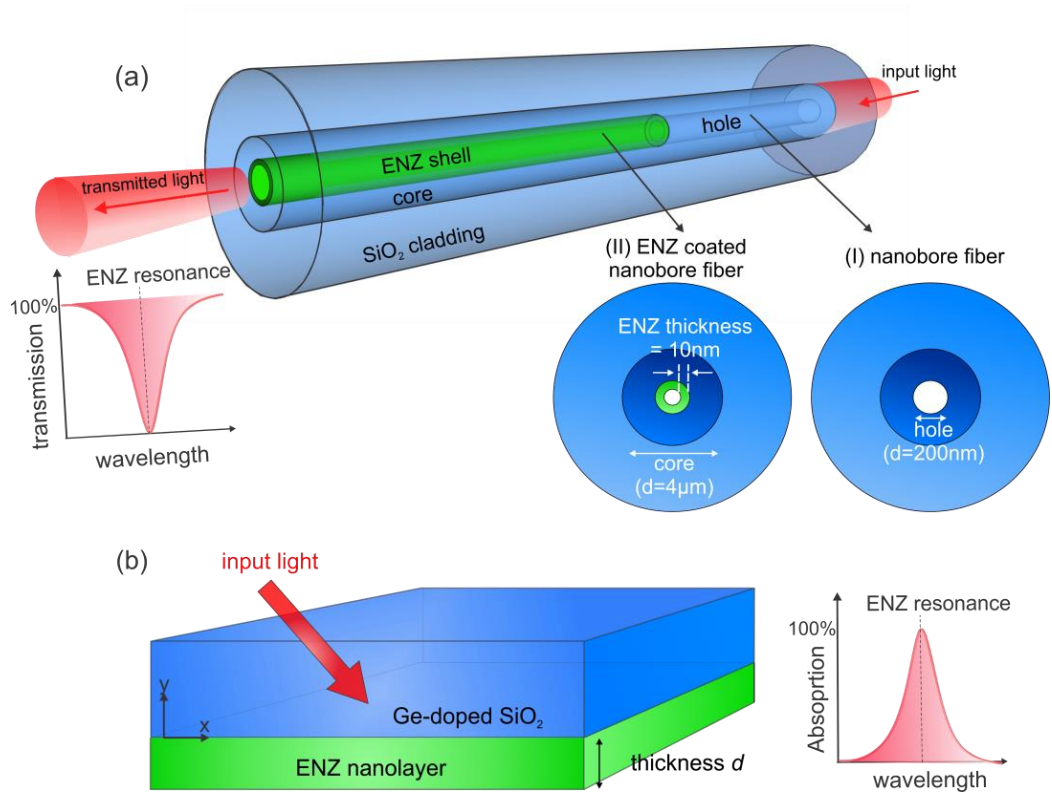

Figure 1. (a) Schematic of the proposed ENZ fiber waveguide design. The inserts are cross-sections of the ENZ fiber (coated with ITO ENZ nano-shell) and hollow nanobore fiber (without ITO ENZ nano-shell). (b) Geometry of glass (GeO$_2$ doped silica)-ITO-air three layer structure with ITO layer thickness $d$ for excitation of NR-ENZ thin film mode.

.

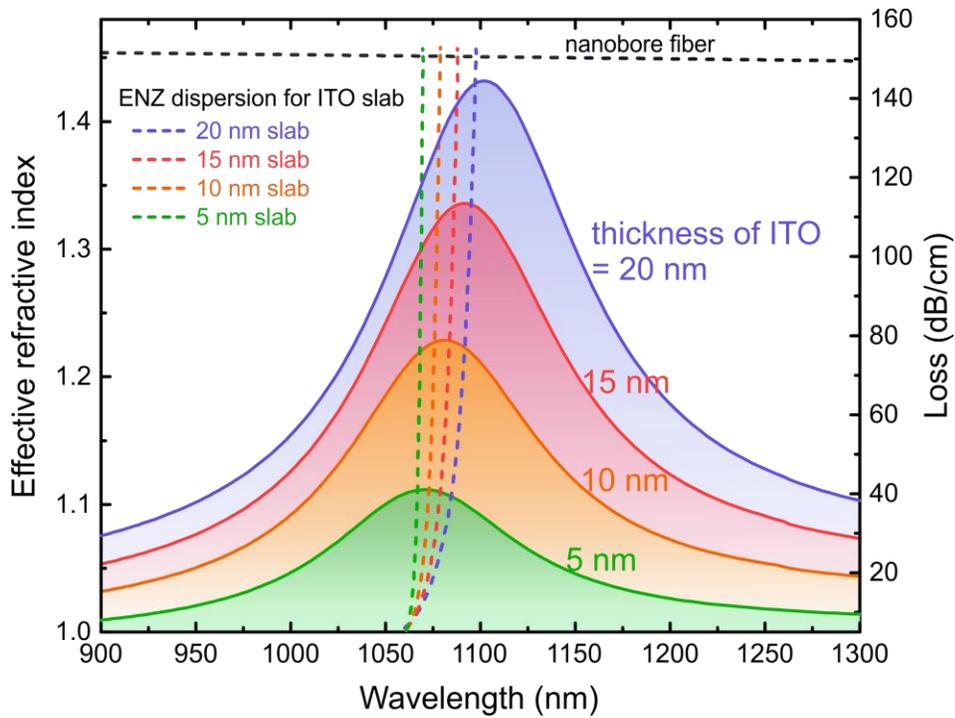

Figure 2. Phase matching conditions between fundamental waveguide mode in the hollow nanobore fiber and thin-film NR-ENZ mode of glass-ITO-air three-layer geometry for 5 nm, 10 nm, 15 nm and 20 nm ITO layer thicknesses. The dotted curves are effective refractive index of waveguide and thin-film modes as a function of wavelength. The solid curves and color shaded areas are modal loss spectra of fundamental mode excited in the ENZ fiber for four ITO shell thicknesses described above.

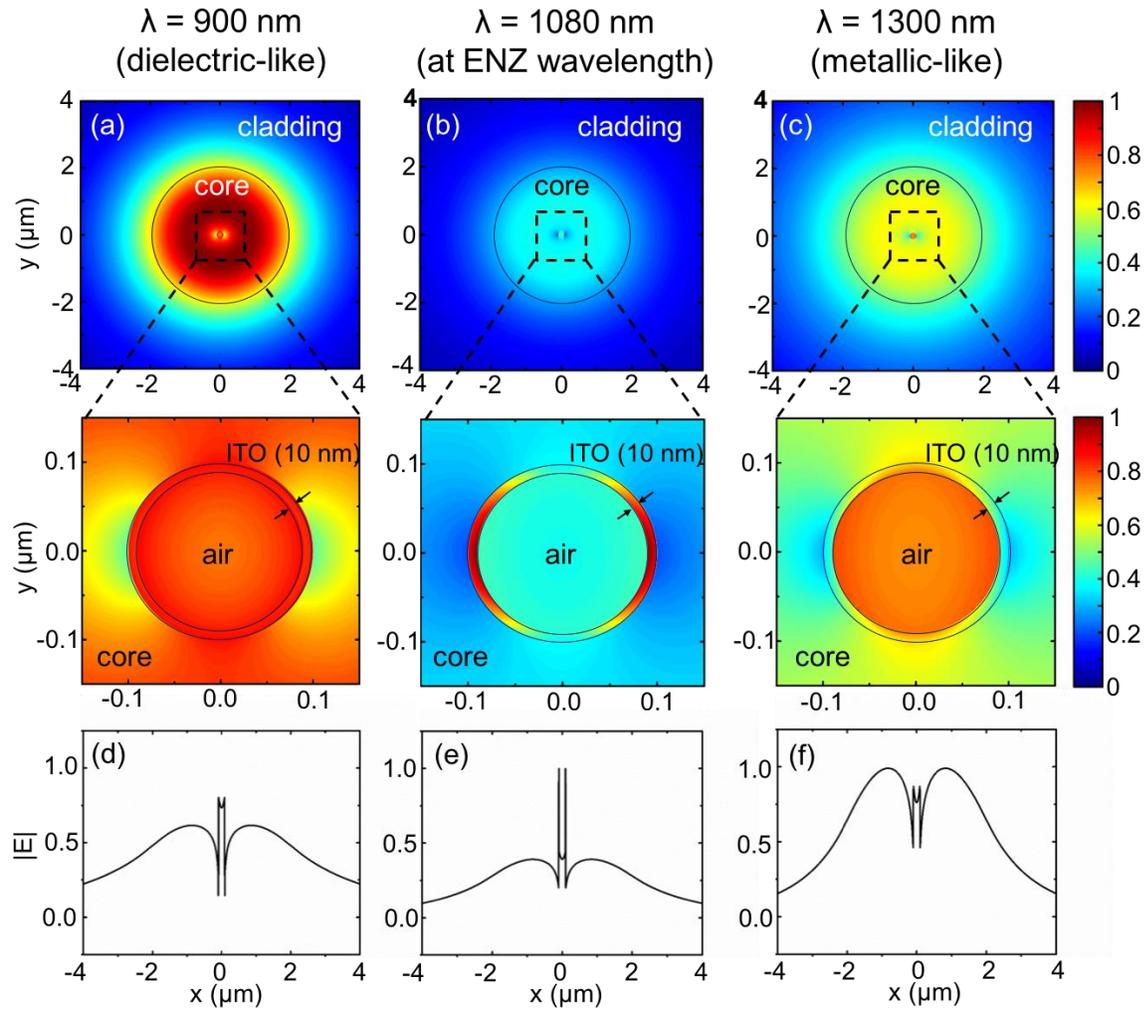

Figure 3. Electric field profile of the fundamental mode supported by the ENZ fiber with 10nm ITO shell thickness at wavelengths (a, d) 900 nm (outside ENZ wavelength), (b, e) 1080 nm (at the ENZ wavelength), (c, f) 1300 nm (outside ENZ wavelength). In (a-c), the top images are the contour plots of electric field magnitude on the entire fiber cross-section and the bottom images zoom in on the ITO nanoshell. The black circles outline the structure of the fiber. (d-f) depict |E| along the diameter of the fiber.

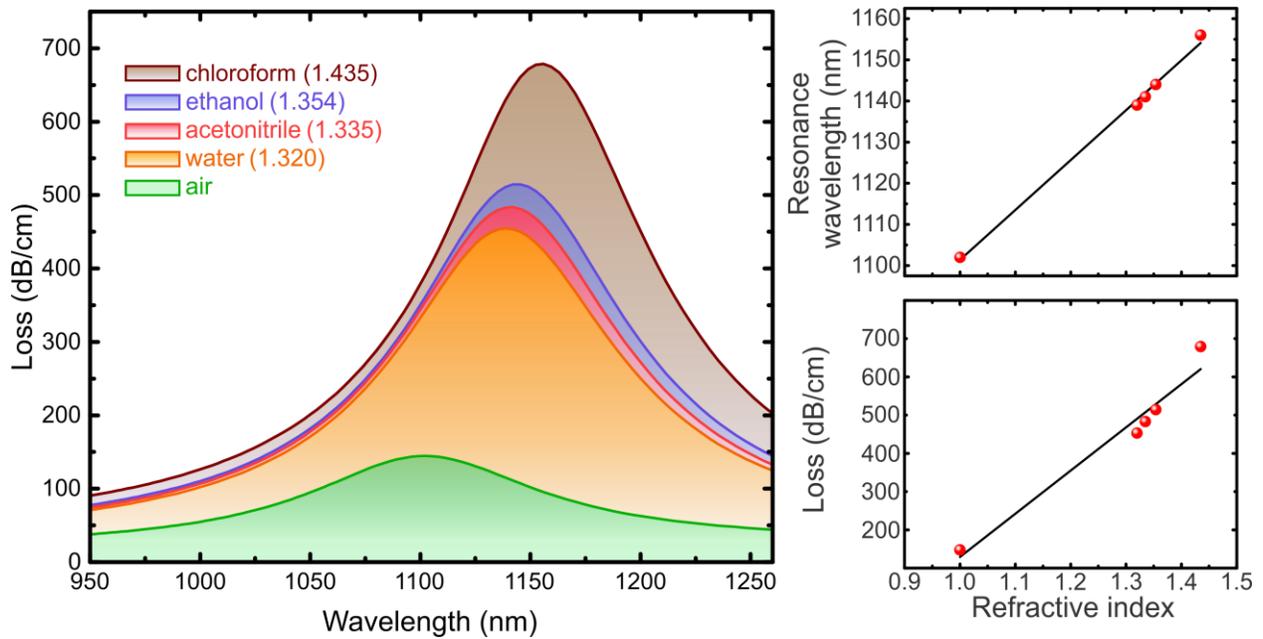

Figure 4. Refractive index dependence of the ENZ fiber. The insert on the left is a sketch of the proposed index sensing mechanism where the analyte is placed inside the central channel of the fiber. The graph plots the loss spectra of ENZ mode for five analytes whose refractive indices at the resonant wavelengths are given within the parentheses. The two inserts on the right shows the refractive index dependence of ENZ resonance wavelength (slope = 121 nm/RIU) and loss on the top and bottom scatter plots respectively. The red lines in the inserts are the linear fits to the data points obtained from the main graph.

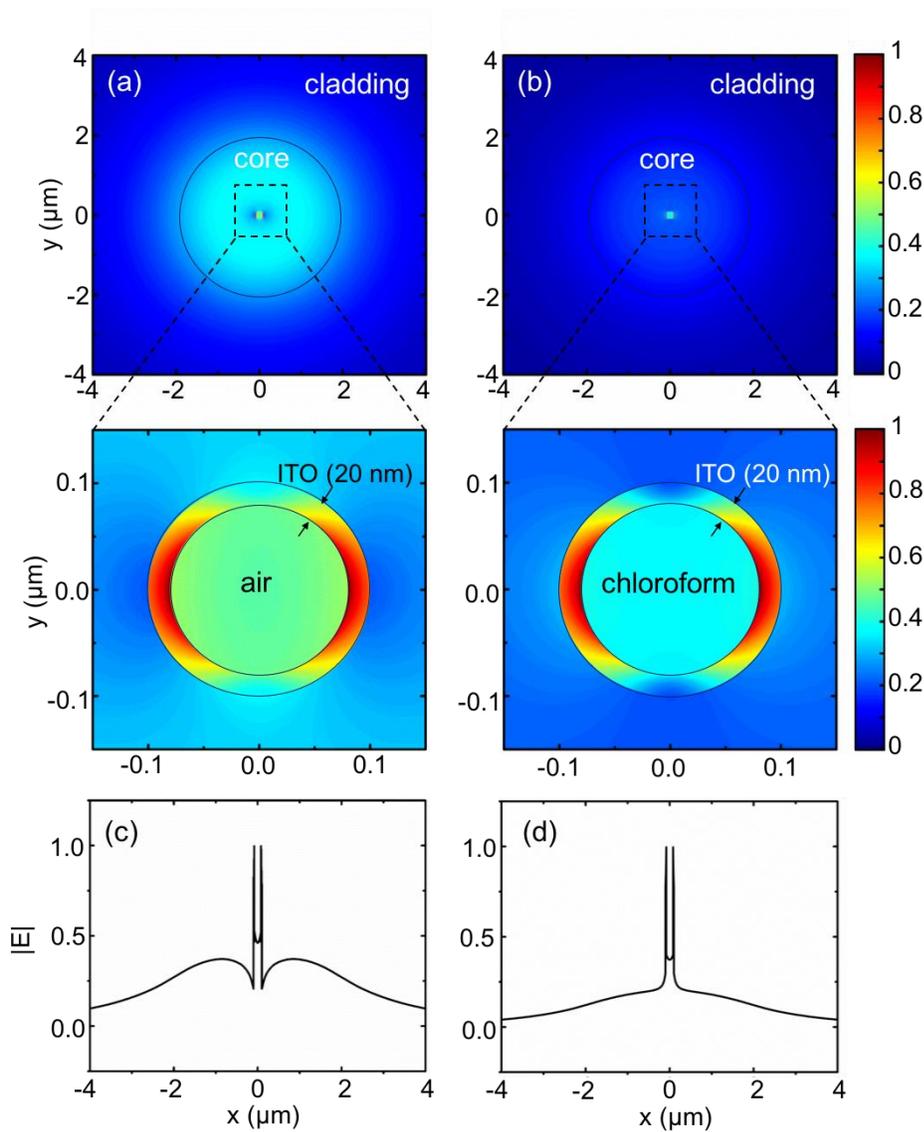

Figure 5. Electric field profiles at ENZ resonance for (a,c) air, and (b,d) chloroform as analytes placed inside the ENZ fiber with 20 nm thick ITO shell. In (a) and (b), the top images are the contour plots of |E| on the entire fiber cross-section and the bottom ones zoom in on the ITO shell. The black circles outline the structure of the fiber. (c) and (d) depict |E| along the diameter of the fiber.